\documentclass[
aps,prb,twocolumn,showpacs,preprintnumbers,amsmath,amssymb,floatfix]{revtex4-2}
\usepackage{graphicx}
\usepackage{subfigure}
\usepackage{dcolumn}
\usepackage{bm}
\usepackage{color}
\usepackage{epstopdf}%
\usepackage{amsmath}%
\usepackage{amsfonts}%
\usepackage{amssymb}
\usepackage{bm}
\newcommand{\mb}{\mathbf}

\begin{document}
\title{Magnetochiral anisotropy-induced nonlinear planar Hall effect in Topological Insulator surface states}
\author{D. C. Marinescu}
\author{S. Tewari}
\affiliation{Department of Physics and Astronomy, Clemson University, Clemson, South Carolina 29634, USA}

\begin{abstract}
In an intriguing recent experiment, it has been found that the two-dimensional (2D) surface states of a three-dimensional (3D) strong topological insulator (TI) support a non-zero Hall voltage transverse to an applied electric field even when the external magnetic field is in the plane (i.e., the in-plane Lorentz force vanishes). This so-called planar Hall effect (PHE) of TI surface states is found to be non-linear, i.e., the Hall voltage scales quadratically with the applied electric field and linearly with the in-plane magnetic field. In this paper, we derive the non-linear PHE for strong topological insulator surface states and show that the derivations contained in the previous literature are incomplete, which can lead to quantitative or even in some cases qualitative errors in the estimates for the nonlinear planar Hall resistance. We derive the complete expressions for the non-linear planar Hall currents for TI surface states with broken particle-hole symmetry and provide results for different regimes of the surface state Hamiltonian
which can be compared with future experiments.
\end{abstract}
\date{\today}
\maketitle
\section{Introduction}
In the conventional Hall effect \cite{Hall, ashcroft}, an electric voltage is generated in a conductor transverse to an applied current in the presence of an out-of-plane magnetic field that deflects the charge carriers in the 2D plane by Lorentz force. In the anomalous  Hall effect the role of the external out-of-plane magnetic field is played by an intrinsic magnetization or non-trivial geometric properties of the band structure such as a non-zero Berry curvature perpendicular to the 2D plane \cite{Luttinger1, Luttinger2, Jungwirth, Xiao}. Such first-order current response to an applied  electric field can be written as $j_{\alpha}=\sigma_{\alpha \beta}E_{\beta}$  ($\alpha,\beta =x,y$) where $\sigma_{\alpha \beta}$ is the conductivity tensor. The conductivity tensor $\sigma_{\alpha \beta}$ can always be decomposed into components that are symmetric and anti-symmetric in the indices $\alpha$ and $\beta$, $\sigma_{\alpha \beta}=\frac{1}{2}(\sigma^s_{\alpha \beta}+\sigma^a_{\alpha \beta})$. The antisymmetric part of the conductivity tensor, $\sigma^{a}_{\alpha \beta}=-\sigma^{a}_{\beta \alpha}$, can be written as, $\sigma^a_{\alpha\beta}=\gamma B_z\epsilon_{\alpha\beta}$, where $\gamma$ is a constant, $\epsilon_{\alpha\beta}$ is the 2D Levi-Civita tensor ($\epsilon_{\alpha\beta}=-\epsilon_{\beta\alpha}$) and $\vec{B}=B_z\hat{z}$ is an external magnetic field. This leads to, $j_{\alpha}=\sigma^s_{\alpha \beta}E_{\beta} + \gamma\epsilon_{\alpha \beta} B_z E_{\beta}=\sigma^s_{\alpha \beta}E_{\beta} - \gamma(\vec{E}\times (B_z\hat{z}))_{\alpha}$, where the magnetic field $\vec{B}$ is assumed to point in the $\hat{z}$-direction and  $\sigma^s_{\alpha \beta}=\sigma^s_{\beta\alpha}$ is the symmetric part of the conductivity tensor. Since the Joule heating is proportional to $\vec{j}\cdot\vec{E}$, it is clear that the symmetric part $\sigma^s$ leads to non-zero Joule heating, while the antisymmetric part $\sigma^a$ leads to zero Joule heating (i.e., $\vec{E}\cdot(\vec{E}\times (B_z\hat{z}))=0$) and is thus non-dissipative. Since the current density $j_\alpha$ is odd under time-reversal symmetry (TRS) while the electric field is even, and they are related in the first order of the electric field by the formula $j_{\alpha}=\sigma_{\alpha \beta}E_{\beta}$, the conductivity $\sigma_{\alpha \beta}$ must break the time-reversal symmetry. The dissipative part of the conductivity tensor $\sigma^s_{\alpha \beta}$ breaks TRS because it leads to Joule heating which is irreversible in time. But, the antisymmetric part of the conductivity tensor $\sigma^a_{\alpha \beta}$ is non-dissipative, implying that the time-reversal breaking in the systems with a non-zero $\sigma^a$ must be intrinsic. The Hall effect in the linear order of the electric field, defined as the non-dissipative, anti-symmetric component of the conductivity tensor, thus necessarily requires broken time-reversal symmetry \cite{Philippe, Onsager1, Onsager2}. Theoretical and experimental studies of the Hall effect in the linear response regime (linear order in the external electric field) have led to important advances in condensed matter physics leading to the search for topological phases of matter \cite{Klitzing, Thouless, Tsui, Hasan, Qi}.

In the nonlinear response regime ($j_{\alpha}=\chi_{\alpha\beta\gamma}E_{\beta}{E}_{\gamma}$), a non-zero Hall-like transverse voltage quadratically dependent on the applied electric field can result with or without broken TRS \cite{Spivak, Sodemann}. In time-reversal symmetric systems with broken space inversion (SI) symmetry, the nondissipative nonlinear Hall current can result from a non-zero first-order moment of the Berry curvature, the so-called Berry curvature dipole (BCD), over the occupied bands \cite{Sodemann}. Such BCD-induced nonlinear Hall effect in TR symmetric systems has recently attracted immense attention in the literature \cite{Kang, Du}. But when TRS and SI are simultaneously broken, a second class of nonlinear Hall-like response is allowed by symmetry, arising from magnetochiral anisotropy effect also known as nonreciprocal magnetotransport \cite{Tokura2018, Yokouchi, Yasuda}. Nonreciprocal magnetotransport implies that in systems with broken SI symmetry subjected to a magnetic field, the resistivity parallel and perpendicular to the applied current is different for the current flowing to the right ($+I$) and left ($-I$) directions.
\begin{figure}
  \centering
 \includegraphics[width=8 cm, height = 5cm,keepaspectratio]{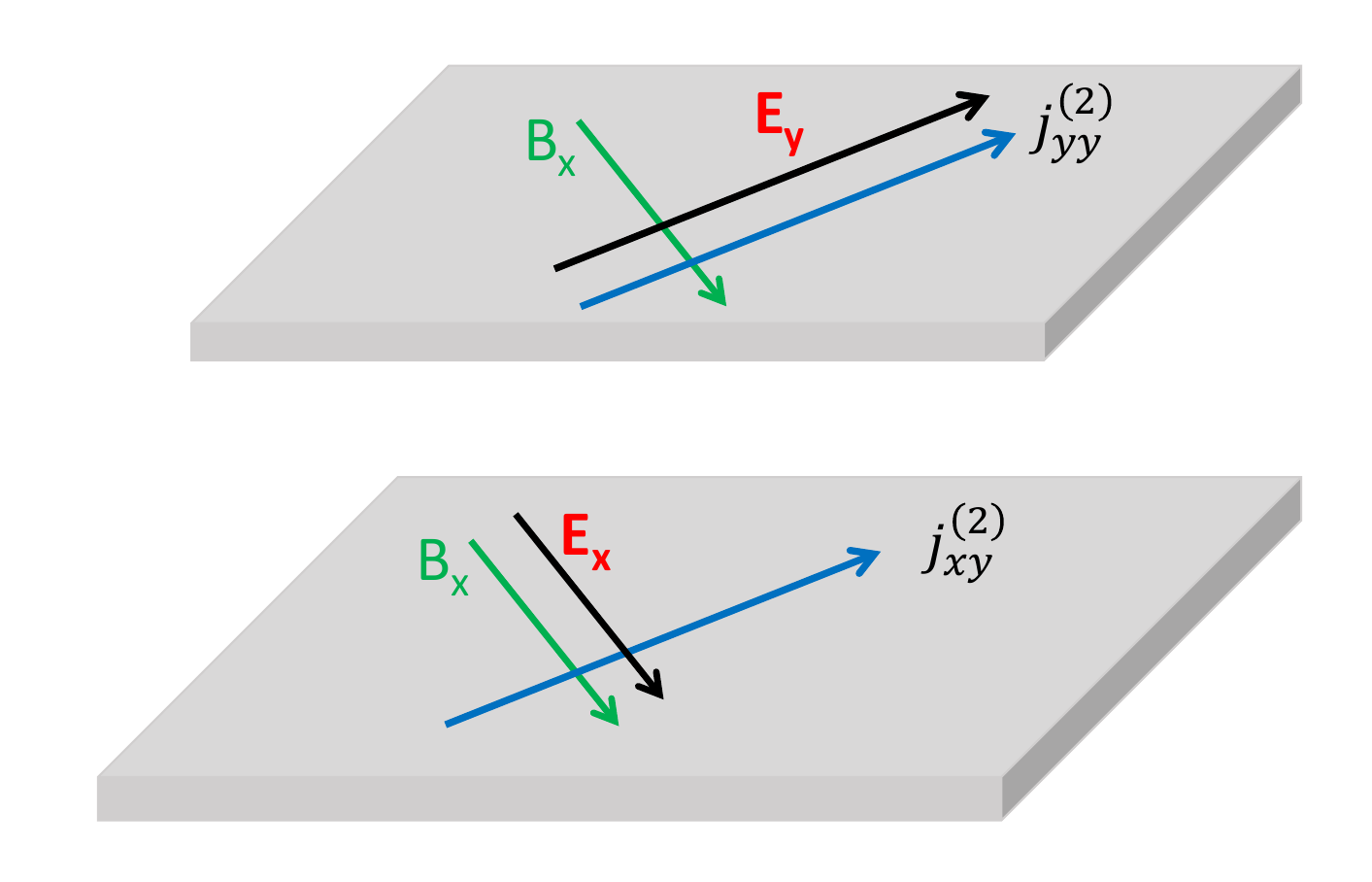}
  \caption{The experimental setup for determining the longitudinal magnetoresistance, when the electric and magnetic fields are perpendicular to each other (top) and the planar Hall effect, when the electric and magnetic fields are parallel (bottom). }
  \label{fig0}
\end{figure}

The nonreciprocal longitudinal magnetoresistance appears when the electric and in-plane magnetic fields are perpendicular to each other and the change in conductivity due to the magnetic field is measured parallel to the applied electric field as shown in Fig.~\ref{fig0}, top panel. Thus, for a magnetic field along the $\hat{x}$ axis, $B_x$, and an electric field along the $\hat{y}$ axis $E_y$, the resistance associated with the non-reciprocal effect is $R=R^0(1+\gamma_{yL} B_xI_y)$ ($R^0$ is the longitudinal linear resistance of the system)
with $\gamma_{yL}$, the longitudinal coupling parameter, given by
\begin{equation}
 \gamma_{yL} = -\frac{1}{AB_x}\frac{\sigma_{2yy}}{(\sigma_{1})^2} \;,\label{eq:gamma-x}
 \end{equation}
 for a system with cross section area $A$. Eq.~(\ref{eq:gamma-x}) is obtained from Ohm's law written for a current density $j_y= \sigma_1E_y + \sigma_{2yy}E_y^2$, where $\sigma_{2yy}$ and $\sigma_1$ and the second order and the linear conductivities, respectively.

 A similar formula can be written down for the Hall resistance, when the electric and magnetic fields are parallel, as in Fig.~\ref{fig0}, bottom panel, and the resistance is measured perpendicular to the applied electric field. Then, $R_H=R_H^0(1+\gamma_{yH} B_xI_x)$, which follows from a nonlinear Hall voltage $V_y=R_H^0 I_x + R_H^{(2)}I_x^2$, where $R_H^0$ could be a possible linear order planar Hall resistance and the nonlinear planar Hall resistance $R_H^{(2)}$ depends linearly on the in-plane magnetic field $B_x$. In this case, the planar-Hall coupling parameter is
 \begin{equation}
 \gamma_{yH} = -\frac{1}{AB_x}\frac{\sigma_{2xy}}{(\sigma_{1})^2}\;.\label{eq:gamma-y}
 \end{equation}

The magnetochiral anisotropy-induced nonlinear magnetoresistance effect \textit{parallel} to the applied electric field when the applied $\mathbf{E}$ and $\mathbf{B}$ fields are perpendicular was first predicted in Rashba semiconductors and was experimentally observed \cite{Rikken2005, Ideue2017, Li2021, Nagaosa2018, Pan2022}. The analogous magnetochiral anisotropy-induced nonlinear Hall effect \textit{transverse} to the applied electric field and in the presence of an \textit{in-plane} parallel magnetic field, the so-called nonlinear planar Hall effect,
was found to be non-zero in Weyl semimetals \cite{Yokouchi}, topological insulator surface states \cite{Yasuda, He_PRL_2019}, and in theoretical calculations in Rashba systems \cite{Marinescu_2023} and systems with cubic Dresselhaus interactions \cite{Dantas_PRB_2023}.


In recent experiments, it was found that the surface states of a 3D strong topological insulator (TI) support a non-zero Hall voltage transverse to an applied electric field in the presence of an \textit{in-plane} magnetic field, i.e., the so-called planar Hall effect (PHE) which scales quadratically with the applied electric field and linearly with the in-plane magnetic field. In this paper, we derive and analyze the non-linear PHE for the surface states of strong TI and show that the derivations contained in the previous literature are at best incomplete, which can lead to quantitative or even in some cases qualitative errors while estimating the expected values of the nonlinear planar Hall resistance. We derive the complete expressions for the non-linear planar Hall currents for TI surface states with broken particle-hole symmetry and provide complete results for different regimes of the surface state Hamiltonian. Our results can be compared with future experiments on the nonlinear planar Hall effect in TI surface states in different regimes of the surface state Hamiltonian which can be accessed by tuning the chemical potential.

In the remainder of this paper, in Section II we introduce the Hamiltonian of TI surface states including the hexagonal warping interaction and a Zeeman energy term, and discuss the energy spectrum. We discuss in detail the different regimes of this Hamiltonian (Dirac fermion regime and 2D electron gas regime with an effective spin-orbit interaction) in terms of an interpolation parameter $t$. In Section III, we discuss the Boltzmann transport formalism in the nonlinear regime used in this paper to derive the second-order Hall-like conductivities quadratic in the applied electric field $E_x$. In Section IV, we present the main results of this paper, namely, the second-order transverse current, $j_{xy}^{(2)}$, for a general value of the interpolation parameter $t$, and use it to calculate the second-order Hall-like conductivities in the various regimes of the Hamiltonian given in Eq.~(\ref{rashba-2D}).  In Section V, for the sake of completeness, we calculate
the magnetochiral anisotropy-induced nonlinear magnetoresistance in TI surface states in a geometry in which the electric and the
in-plane magnetic fields are perpendicular to each other. In this geometry, the change in resistance is measured parallel to the
electric field, while in the nonlinear planar Hall effect the resistance is measured in a direction transverse to the electric field.  We end with a summary and conclusion in Section VI.

\section{Hamiltonian of TI Surface States and the Energy Spectrum}

The generic Hamiltonian of an electron of momentum $\mathbf{p} = \{p_x,p_y\}$, spin $\boldsymbol{\sigma} = \{\sigma_x,\sigma_y,\sigma_z\}$, and effective mass $m^*$ on the 2D surface states of 3D time-reversal symmetric strong topological insulators is given by \cite{Phillips, Fu_Warping, Liu, McKenzie_2013},
\begin{equation}
H_{TI}=\frac{p_x^2 + p_y^2}{2m^*} + \alpha (\sigma_x p_y - \sigma_y p_x)\;.
\label{eq:HTI}
\end{equation}
Here, $\alpha$ is the coupling constant representing the effective velocity of the Dirac fermion part of the Hamiltonian. We will use $\epsilon_F$ for the Fermi energy (chemical potential) in the TI surface measured from the band crossing point.
The first term in $H_{TI}$ quadratic in the momentum and proportional to the inverse effective mass breaks particle-hole and sublattice symmetries \cite{Schnyder}. This term is always present in the surface states of real topological insulators. $H_{TI}$ describes pure Dirac Fermions for $m^* \rightarrow \infty$ and a normal 2D electron gas for $\alpha=0$. The transition between the two limits of the Hamiltonian can be described by an interpolation parameter $t$ given by,
\begin{equation}
t = \frac{m^*\alpha}{\sqrt{(m^*\alpha)^2 + 2m^*\epsilon_F}} \;.
\label{eq:t}
\end{equation}
For $2m^*\epsilon_F \gg (m^*\alpha)^2$ (i.e., $\epsilon_F\gg m^*\alpha^2/2$), the interpolation parameter $t$ approaches, $t\sim\frac{m^*\alpha}{\sqrt{2m^*\epsilon_F}}\rightarrow 0$. In this limit, the Hamiltonian in Eq.~(\ref{eq:HTI}) is that of a 2DEG with a spin-orbit interaction (SOI) with the effective SOI coupling constant given by the Dirac velocity $\alpha$.
By contrast, if $m^* \rightarrow \infty$, $2m^*\epsilon_F \ll (m^*\alpha)^2$ (i.e., $\epsilon_F\ll m^*\alpha^2/2$) and $t \rightarrow 1$, so the interpolation parameter $t=1$ describes pure Dirac fermions. For $1 > t > 0$, the system behaves as a 2DEG with an effective spin-orbit coupling given by the Dirac velocity $\alpha$, while $t \sim 1$ implies a graphene-like Dirac fermion system.

Note also that the Hamiltonian in Eq.~(\ref{eq:HTI}) produces a pair of distinct Fermi surfaces for all band fillings. However, in real materials, $H_{TI}$ is only an effective low-energy Hamiltonian valid near the zone center and away from the first Brillouin zone boundary, and therefore
the
outer Fermi surface may not actually exist. When both bands and both Fermi surfaces are considered, Eq.~(\ref{eq:HTI}) formally corresponds to the Hamiltonian of a 2DEG with a Rashba spin-orbit coupling.

In the presence of a magnetic field aligned along the $\hat{x}$ axis, $B_x$, and a hexagonal warping effect $\tilde{\lambda}$ \cite{Fu_Warping}, Eq.~(\ref{eq:HTI}) becomes,
\begin{equation}
H_{2D} = \frac{p_x^2+p_y^2}{2m^*} + \alpha(p_y\sigma_x-p_x\sigma_y)+B_x\sigma_x + \tilde{\lambda}\sigma_z\;.\label{rashba-2D}
\end{equation}
For simplicity, in Eq.~(\ref{rashba-2D}), $B_{x}$ designates the Zeeman splitting associated with the projections of the magnetic field along the $\hat{x}$ direction $g\mu_B B_{x}\sigma_x/2$ with $\mu_B$ the Bohr magneton and $g$ the effective gyromagnetic factor. We have included the magnetic field in the $x$-direction in anticipation of our calculation below of a non-zero transverse Hall-like current in the $y$-direction proportional to the first power of $B_x$ and second power of an applied electric field $E_x$ (see Eq.~(\ref{eq:TI-current*})). In Eq.~(\ref{rashba-2D}) the warping interaction $\tilde{\lambda}$ is cubic in the magnitude of the electron momentum, $p = \sqrt{p_x^2 + p_y^2}$, and is given by,
\begin{equation}
\tilde{\lambda} = \lambda p^3\cos3\varphi = \lambda (p_x^3 - 3p_xp_y^2)
\label{eq:warping}
\end{equation}
where $\tan\varphi = p_y/p_x$ and $\lambda$ is the coupling constant.

\begin{figure}
    \centering
  {\includegraphics[width=8 cm, height = 5cm,keepaspectratio]{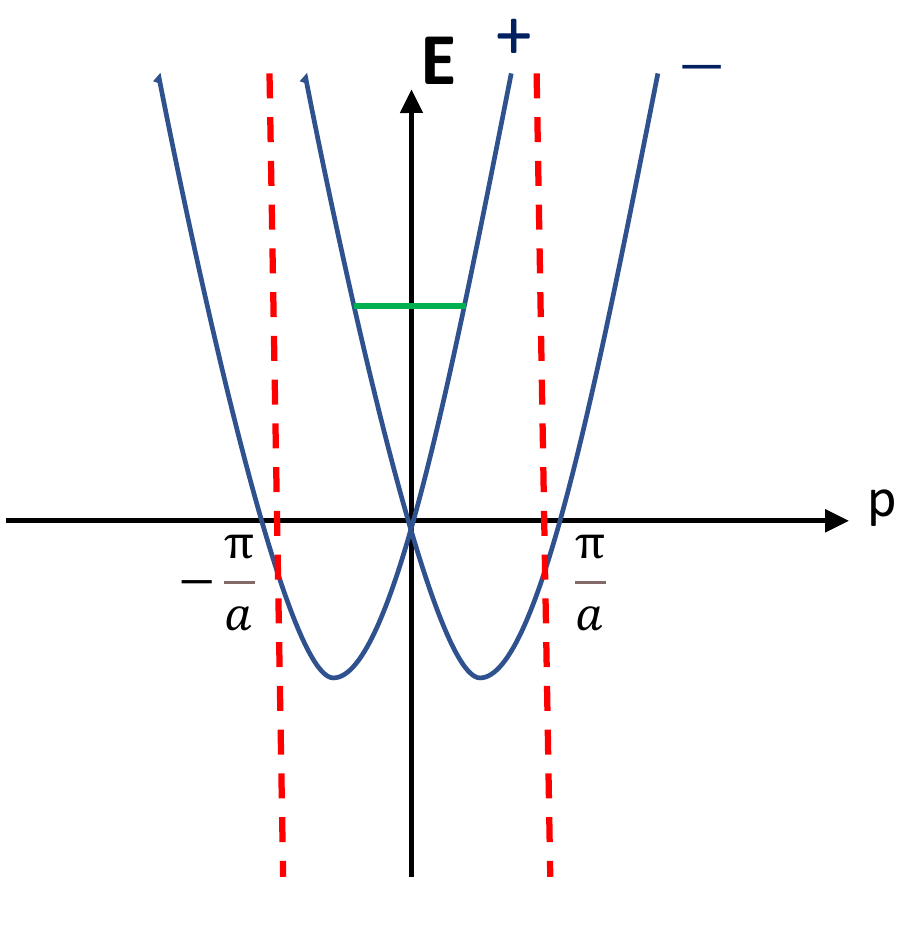}}
    \caption{ The energy spectrum of a topological insulator with finite mass $m^*$ in (b). $+$ and $-$ designate the two chiral eigenvalues. The dashed vertical lines represent the Brillouin zone limits. The second Fermi surface is excluded from the calculation since the corresponding Fermi momentum is assumed to be larger than the limit of the Brillouin zone. The Fermi energy is the horizontal line. In this regime, the surface state of a 3D topological insulator is a single-band system. }
    \label{fig1}
\end{figure}

The eigenvalues of the Hamiltonian, designated by the chiral index $\xi = \pm 1$, are
\begin{equation}\label{eigenvalues-zero}
  E_\xi = \frac{p^2}{2m^*} + \xi \sqrt{\alpha^2\left[p_x^2 + \left(p_y + \frac{B_x}{\alpha}\right)^2\right] + \tilde{\lambda}^2}\;.
\end{equation}
The energy spectrum in the absence of a magnetic field and warping is presented in Fig.~\ref{fig1} .
The following considerations assume that the magnetic term $B_x$ and the warping term $\tilde{\lambda}$ are {much smaller than the Dirac energy $\alpha p$ for a given momentum $p$. It is clear from Eq.~(\ref{eigenvalues-zero}) that the effect of the applied magnetic field is to shift the $p_y$ momentum by a constant value $-B_x/\alpha$, while the warping is present in this problem only as $\tilde{\lambda}^2$. Thus, the lowest order effects that reflect these changes in the electron energy are linear in the magnetic field and quadratic in $\lambda$. }

With this insight, we write
\begin{equation}
p_x^2 + \left(p_y + \frac{B_x}{\alpha}\right)^2 = \left(p + \frac{B_x\sin\varphi}{\alpha}\right)^2 + \frac{B_x^2\cos^2\varphi}{\alpha^2}\;,
\end{equation}
and approximate
\begin{eqnarray}\label{eigenvalues}
&& E_\xi = \frac{p^2}{2m^*} + \xi \alpha \left(p + \frac{B_x\sin\varphi}{\alpha}\right)\nonumber\\
&&\times  \left[1+\frac{\tilde{\lambda}^2 + \frac{B_x^2\cos^2\varphi}{\alpha^2}}{2\alpha^2 \left(p + \frac{B_x\sin\varphi}{\alpha}\right)^2}\right]\;.
\end{eqnarray}
Henceforth, in Eq.~(\ref{eigenvalues}), terms quadratic in $B_x$ will be neglected.
By introducing function $W$,
\begin{equation}\label{eq:w}
W = \frac{\tilde{\lambda}^2}{2\alpha \left(p + \frac{B_x}{\alpha}\sin\varphi\right)}\;,
\end{equation}
we arrive at an expression of the eigenvalues that depends only on $B_x$ and $W$,
\begin{eqnarray}\label{exi}
  &&E_\xi = \frac{p^2}{2m^*} + \alpha p + \xi (B_x\sin\varphi + W) \nonumber\\
  && = \frac{\left(p +\xi m^*\alpha\right)^2}{2m^*}+ \xi (B_x\sin\varphi + W) -\frac{(m^*\alpha)^2}{2m^*}
\end{eqnarray}

In the following discussions, we assume that the Fermi energy is positive (as measured from the band crossing point, see Fig.~(\ref{fig1})), $\epsilon_F>0$. {The condition $\epsilon_F = E_\xi$  generates the Fermi momenta associated with this system,}
\begin{eqnarray}\label{pxi}
  p_\xi &=& \sqrt{2m^*[\epsilon_F - \xi (B_x\sin\varphi + W)] + (m^*\alpha)^2} \nonumber\\
  &&- \xi m^*\alpha\;.
\end{eqnarray}
As is the case for a Rashba system, there are two Fermi surfaces corresponding to the two chiral bands $\xi=\pm 1$ and two associated Fermi momenta $p_{+}$ and $p_{-}$. However, for topological insulators, we will assume that the outer Fermi surface does not exist because the corresponding Fermi momentum is larger than a momentum cutoff, which we take to be the first Brillouin zone boundary, below which the Hamiltonian in Eq.~(\ref{rashba-2D}) is supposed to hold.

Since the warping contribution $W$ depends explicitly on the momentum (see Eq.~(\ref{eq:warping})) we evaluate it at the Fermi momenta for $W=0$ given by,
\begin{equation}\label{pxi2}
({p}_{\xi})_{W = 0} = \sqrt{2m^*(\epsilon_F - \xi B_x\sin\varphi) + (m^*\alpha)^2} - \xi m^*\alpha \;.
\end{equation}
Thus, the value of $W$ depends on $\xi$ and $B_x$.

The analysis of the Fermi momenta in the absence of warping and magnetic field provides the physical justification of the interpolation parameter $t$. We first write,
\begin{equation}
({p}_{\xi})_{B_x = 0, W = 0} = \sqrt{2m^*\epsilon_F  + (m^*\alpha)^2} - \xi m^*\alpha\;.
\label{eq:pxi3}
 \end{equation}
  In the Dirac fermion limit ($\epsilon_F\ll m^*\alpha^2/2$, i.e., $t\rightarrow 1$) we obtain the Fermi momentum for $\xi=+1$,
 \begin{equation}
 p_1 = \sqrt{2m^*\epsilon_F  + (m^*\alpha)^2} - m^*\alpha = m^*\alpha \frac{1-t}{t}\;.\label{eq:p1*}
 \end{equation}
  Since in this limit $t$ approaches $1$ as $t = 1-\frac{\epsilon_F}{m^*\alpha^2}$, Eq.~(\ref{eq:p1*}) gives the dispersion relation of a topological insulator surface state, $p_1\rightarrow \epsilon_F/\alpha$.
At the same time, in this limit for $\xi = -1$, we find,
\begin{equation}\label{eq:p-1}
p_{-1} = \sqrt{2m^*\epsilon_F  + (m^*\alpha)^2} + m^*\alpha = m^*\alpha\frac{1+t}{t}\;,
\end{equation}
so, for $t\rightarrow 1$, $p_{-1} = 2m^*\alpha + \frac{\epsilon_F}{\alpha}\simeq 2m^*\alpha$.

Therefore, in the Dirac fermion limit ($t\rightarrow 1$), $p_{-1}\simeq 2m^*\alpha$ becomes larger than the Brillouin zone boundary when $2m^*\alpha \gtrsim \frac{\pi}{a}$, leading to a constraint on the effective mass that eliminates the outer Fermi surface. 
{More generally, the condition on the effective mass $m^*\gtrsim \frac{\Lambda}{2\alpha}$, where $\Lambda$ is a wavenumber cutoff depending on the microscopic parameters of the system (below which the Hamiltonian in Eq.~(\ref{rashba-2D}) is valid) and $\alpha$ is the Dirac velocity in Eq.~(\ref{rashba-2D}), effectively eliminates the $\xi = -1$ chiral band from transport considerations. We will assume this condition naturally holds for topological insulators with values of the interpolation parameter $t \sim 1$.
To preserve this regime of single-band transport for a general value of the interpolation parameter $0< t <1$, one needs to use Eq.~(\ref{eq:p-1}) to establish the relationship between $\epsilon_F$ and $m^*\alpha$ that allows it for a given $\Lambda$, while still satisfying the constraint that $\epsilon_F< m^*\alpha^2/2$ so that the dispersion relation for a topological insulator is recovered for $p_1$.

In the limit $t \rightarrow 1$, Eq.~(\ref{eq:pxi3}) generates, in a second order expansion,
\begin{equation}
p_\xi = m^*\alpha \left[1+\frac{2m^*\epsilon_F}
{2(m^*\alpha)^2} - \frac{1}
{8}\left(\frac{2m^*\epsilon_F}
 {(m^*\alpha)^2}\right)^2\right] - \xi m^*\alpha;.
\end{equation}
For $\xi = 1$, we obtain,
\begin{equation}
p_1 = \frac{\epsilon_F}{\alpha}\left[1 - \frac{\epsilon_F}{2m^*\alpha^2}\right]\;,
\end{equation}
which is the Fermi momentum of a topological insulator corrected for the finite mass $m^*$. We note that this approximation is valid only for  $\epsilon_F\ll m^*\alpha^2/2$.

In the other limit of the interpolation parameter $t\rightarrow 0$, Eq.~(\ref{eq:pxi3}) generates $p_\xi \simeq \sqrt{2m^*\epsilon_F}(1-\xi t)$. This corresponds to $m^*\alpha^2/2\ll \epsilon_F$, a regime where the two momenta are almost equal to $\sqrt{2m^*\epsilon_F}$, which is the Fermi momentum of a 2DEG. Therefore, for $t\rightarrow 0$, when
the Hamiltonian in Eq.~(\ref{eq:HTI}) resembles that of a 2DEG with
an effective Rashba-like spin-orbit coupling, both chiral
bands with $\xi = \pm 1$ should contribute to transport.

To summarize this section, using the Hamiltonian $H_{2D}$ in Eq.~(\ref{rashba-2D}), one can obtain both the Dirac fermion regime and single-band transport assuming that for a given $\alpha$ the effective mass and the Fermi energy satisfy certain conditions, and the 2DEG regime with an effective spin-orbit coupling given by the Dirac velocity $\alpha$. In transport, the difference between the two situations is reflected in how many chiral bands are included: only one ($\xi = 1$) for the Dirac fermion regime and both $\xi = \pm 1$ for a 2DEG with effective spin-orbit interaction.

\section{Boltzmann transport formalism in nonlinear regime}

In the presence of an electric field, the electron distribution function $f_\mathbf{p}$
is written as $f_\mathbf{p} = f_\mathbf{p}^0 + \delta f_\mathbf{p}^{(1)} + \delta f_\mathbf{p}^{(2)}$, where $f^0(\epsilon_\mathbf{p}) = \left[e^{(\epsilon_{\mathbf{p}}-\epsilon_F)/k_BT} + 1\right]^{-1}$ is the equilibrium Fermi function, while $\delta f_\mathbf{p}^{(1)}$ and $\delta f_\mathbf{p}^{(2)}$ are the non-equilibrium corrections proportional to $E_x$ and $E_x^2$ respectively.
$\delta f_\mathbf{p}^{(1)}$ is the solution to the Boltzmann transport equation (BTE) in the relaxation time approximation \cite{ashcroft},
\begin{equation}
\delta f_{\mathbf p}^{(1)} = e\tau\mathbf{v}_{\mathbf{p}}\cdot \mathbf E \frac{df^0}{d\epsilon} \;,\label{eq:ef1}
\end{equation}
where $v_{\mathbf{p}} = \nabla_{\mathbf p}\epsilon_\mathbf{p}$ is the electron velocity and $\tau$ is the relaxation time.

To obtain the second order distribution function, we use a semi-classical approximation of the local perturbation induced by the external fields on the distribution function \cite{pan}.
 Thus, the addition of an electrostatic potential $V(\mb r) = -\mb E\cdot \mb r$, modifies locally the electron energy to $\tilde{\epsilon}_{\mb p} = \epsilon_{\mb p} + e\mb E\cdot\mb r$, a change considered weak with respect to the Fermi energy and valid within a region $\mb r = v_{\mathbf{p}}\tau$.
 By performing a Taylor expansion in $\left(e\tau\mb E\cdot\mb v_{\mb p}\right)$, the second order correction to the distribution function is obtained to be,
\begin{equation}
\delta f_{\mb p}^{(2)} = \frac{1}{2k_BT}\left(e\tau\mb E\cdot\mb v_{\mb p}\right)^2\tanh \frac{\epsilon_{\mb p} - \epsilon_F}{2k_BT}\left(-\frac{\partial f_{\mb p}^0}{\partial \epsilon_{\mb p}}\right)\;. \label{eq:ef2}
\end{equation}

 {As the shift of the electron momentum along the $\hat{y}$ induced by the in-plane magnetic field is proportional with $B_x/\alpha$, an even function in phase space, a charge current that responds to this perturbation as well as to an electric field is generated only by an even distribution function, namely $\delta f_{\mb p}^{(2)}$ in Eq.~(\ref{eq:ef2}), leading to
\begin{equation}\label{current-0}
j_{xy}^{(2)} = - e\sum_{\mathbf p, \xi}v_{y\xi}\delta f^{(2)}_\mathbf{p}\;.
\end{equation}}
Because $\delta f^{(2)}_\mathbf{p}$ in Eq.~(\ref{eq:ef2}) is proportional to $\tanh[(\epsilon_\mathbf{p} - \epsilon_F)/2k_BT]$ which cancels at $\epsilon_F$, the summation algorithm for the current (Eq.~(\ref{current-0}))  uses the Sommerfeld expansion \cite{ashcroft},
\begin{eqnarray}\label{eq:jy*}
& & j_{xy}^{(2)} = -\frac{e^3 E_x^2}{48\hbar^2}\sum_\xi\tau_\xi^2\int_0^{2\pi}\frac{d}{d\epsilon}\left[p_\xi\frac{dp_\xi}{d\epsilon} v_{y\xi} v_{x\xi}^2\right]_{\epsilon_F}d\varphi\;.\nonumber\\
 \end{eqnarray}

 Although in Eq.~(\ref{eq:jy*}) we sum over the chiral band index $\xi$, as discussed in the last section, in the case of TI surface states, the sum is reduced to $\xi = 1$. {We also note that when conduction involves both bands, the SOI-induced corrections on the relaxation times evaluated at the Fermi energy have to be considered \cite{Verma2019,Kapri2021,pan}. Thus,
for $\epsilon_F>0$,
\begin{equation}
\frac{\hbar}{\tau_\xi} = \frac{\hbar}{\tau}\left(1+\xi\frac{m^*\alpha}{2p_0}\right)\;,\label{eq:tau-xi}
\end{equation}
where $p_0$ is
\begin{equation}
p_0 = \sqrt{(m \alpha)^2 + 2m^* \epsilon_F} \;. \label{eq:p0}
\end{equation}

The electron velocities at the Fermi level, needed in the current calculations, are evaluated in a linear approximation in $B_x$ and $W$,
\begin{eqnarray}\label{vels}
  v_{x\xi} &=& \left(\frac{\partial E_\xi}{\partial p_x}\right)_{\epsilon_F} = \left(\frac{p_\xi + \xi m^*\alpha}{m^*}-\xi\frac{W}{p_\xi}\right)\cos\varphi \nonumber\\
  &&-\xi\frac{B_x\sin\varphi\cos\varphi}{p_\xi}+ \xi\frac{WB_x}{\alpha p^2_\xi}\sin2\varphi\;, \nonumber\\
  &&\\
  v_{y\xi} &=& \left(\frac{\partial E_\xi}{\partial p_x}\right)_{\epsilon_F} =  \left(\frac{p_\xi + \xi m^*\alpha}{m^*}-\xi\frac{W}{p_\xi}\right)\sin\varphi \nonumber\\
  && + \xi\frac{B_x\cos^2\varphi}{p_\xi}-\xi\frac{WB_x}{\alpha p^2_\xi}\cos2\varphi\;,
\end{eqnarray}
There is no direct contribution from warping in the expression of the velocities since the derivatives of the warping interaction are proportional to $\cos3\varphi$ and would disappear under the {angular integration that is involved in the current calculation.}

{\section{Second order transverse currents}}

To simplify the calculation of the current kernel,
we denote by $\bar{K}$ the angular integral of the current expression {in Eq.~(\ref{eq:jy*})},
\begin{eqnarray}\label{eq:K}
  \bar{K}_\xi &=& \int_{0}^{2\pi}v_yv_x^2p_\xi\frac{dp_\xi}{d\epsilon}d\varphi\;,
\end{eqnarray}
{where as functions of the energy the velocities are given by Eq.~(\ref{vels}) and the momenta by Eq.~(\ref{pxi})} written for $\epsilon_F = \epsilon$. Since the warping interaction $W$ present in these expressions is momentum dependent, we make its spatial anisotropy explicit by writing $W = (W_0 + B_x\sin\varphi W_1)\cos^23\varphi$, where as before we keep only the term linear in $B_x$. Thus, from Eq.~(\ref{eq:w}),
$W_0$ and $W_1$ are calculated as
\begin{eqnarray}\label{w-01}
  W_0 &=& \frac{\lambda^2}{2\alpha}{p}_\xi^5 \\
  W_1&=& -\frac{\lambda^2}{2\alpha^2}{p}_\xi^4\left(1+5\xi\frac{m^*\alpha}{p_0}\right)\;.
\end{eqnarray}
$p_\xi$ and $p_0$  are given in Eqs.~(\ref{eq:pxi3}) and (\ref{eq:p0}) respectively written for an energy $\epsilon$ instead of $\epsilon_F$, such that $W_0$ and $W_1$ remain functions of the energy.

In evaluating the density of states ({$p_\xi\frac{dp_\xi}{d\epsilon}$}) from Eq.~(\ref{pxi}), we also have to consider the derivative of $W$ with respect to the energy, $W' = (W_0' + B_x\sin\varphi W_1')\cos^23\varphi$.
{When the kernel in Eq.~(\ref{eq:K}) is linearized in $B_x$, i.e. only linear terms in $B_x$ are considered, and the angular integral is performed, we obtain,}
   \begin{eqnarray}\label{eq:angular-kernel}
  \bar{K}_\xi &=& \frac{\pi}{2}B_x\left[-\xi\left(\frac{p_0}{m^*}\right) -\xi\frac{1}{4} p \left(\frac{p_0}{m^*}\right)^2 W_1' +\frac{1 }{2}\left(\frac{p}{m^*}\right) W_0'\right.\nonumber\\
  &+& \left.\frac{1}{2}\left(\alpha - 3\xi\frac{p_0}{m^*}\right)W_1
 +\frac{W_0}{2 p_0} + \xi \frac{W_0}{2m^*\alpha}\right]\;,
\end{eqnarray}
with $p_\xi$ from Eq.~(\ref{pxi}) reduced to its value in the absence of a magnetic field and warping, $p_\xi  = \sqrt{2m^*\epsilon + (m^*\alpha)^2} - \xi m^*\alpha$.

With input from Eqs.~(\ref{eq:angular-kernel}), (\ref{w-01}), and (\ref{eq:jy*}), we obtain the second order transverse current response $j_{xy}^{(2)}$ proportional to two powers of the applied electric field $E_x$ and linearly dependent on $B_x$ as the derivative of $\bar{K}_\xi$ in respect with the energy evaluated at $\epsilon_F$ as,
\begin{eqnarray}\label{eq:jy}
& & j_{xy}^{(2)} =  \frac{\pi e^3E_x^2 B_x}{96\hbar^2p_0}\sum_{\xi}^{}\xi\tau_\xi^2\left\{1-\frac{1}{8}\left(\frac{\lambda p_\xi^2}{\alpha }\right)^2\left(\frac{p_0}{p_\xi}\right)\right. \nonumber\\
&& \left.\times \left\{40 + 192\xi \left(\frac{m^*\alpha}{p_0}\right) - 40 \left(\frac{m^*\alpha}{p_0}\right)^2\right.\right.\nonumber\\
&& \left.\left. +\left(\frac{p_\xi}{p_0}\right)\left[20 + 35\xi\left(\frac{m^*\alpha}{p_0}\right)+20\left(\frac{m^*\alpha}{p_0}\right)^2\right]\right.\right.\nonumber\\
  & & \left.\left. -7\xi\left(\frac{p_\xi}{p_0}\right)^2\left(\frac{m^*\alpha}{p_0}\right)\right\}\right\}
 \end{eqnarray}
 In terms of the interpolation parameter, $t$, from Eq.~(\ref{eq:t}), we can rewrite the second-order transverse current given in Eq.~(\ref{eq:jy}) as,
 \begin{eqnarray}\label{eq:jy-t}
 & & j_{xy}^{(2)} = \frac{\pi e^3 E_x^2 B_x  }{96\hbar^2 (m^*\alpha)}\sum_{\xi = \pm 1}^{}\xi\tau_\xi^2\left\{t-\frac{1}{8}\left[\frac{\lambda (m^*\alpha)^2}{\alpha}\right]^2(1-\xi t)^3\right.\nonumber\\
 &&\times\left.\left\{40 + 192\xi t - 40 t^2\right.\right.\nonumber\\
&& \left.\left. +\left(1-\xi t\right)\left[20 + 35\xi t+20t^2\right]-7\xi\left(1-\xi t\right)^2t\right\}\right\}
 \end{eqnarray}

Eq.~(\ref{eq:jy-t}) is a general expression for the second-order planar Hall current in a 2D system with Hamiltonian given in Eq.~(\ref{rashba-2D}). Here $\tau_{\xi}$ are the relaxation times for the chiral bands denoted by the chiral index $\xi=\pm 1$ and are given in Eq.~(\ref{eq:tau-xi}).

Below, we discuss its application to the 2D TI surface states in the Dirac fermion limit where we assume that the outer band is excluded from transport. As discussed before the exclusion of the outer Fermi surface is most naturally realized in the limit of the interpolation parameter $t\rightarrow 1$ (see Fig.~(\ref{fig1}), Eq.~(\ref{eq:t}), and the discussion below Eq.~(\ref{pxi2})), but we will provide the results for the second order transverse Hall-like current for 2D topological insulator surface states for a general value of the interpolation parameter $0<t<1$.
For completeness, we will also discuss the second-order planar Hall current for the 2D electron gas regime with an effective Rashba interaction given by the Dirac velocity $\alpha$. In this case, both Fermi surfaces need to be included in the result. As discussed before (see discussion below Eq.~(\ref{pxi2})), this situation is most naturally realized for the interpolation parameter $t\rightarrow 0$. In both cases, it is easy to see that the warping effect enters the second-order current expression through its ratio to the square of the Dirac velocity $\alpha$.

\vspace*{1 cm}
\subsection{2D surface state Hamiltonian in the Dirac fermion regime}

In the case of a topological insulator surface state Hamiltonian in the Dirac fermion regime ($2m^*\epsilon_F\ll (m^*\alpha)^2$, i.e., $t\rightarrow 1$ see Eq.~(\ref{eq:t})), the current is produced only by states with $\xi = +1$ whose relaxation time is taken as $\tau$. Thus,
from Eq.~(\ref{eq:jy-t}) we extract the second order conductivity $\sigma_{2xy} = j^{(2)}_{xy}/E_x^2$,
 \begin{eqnarray}\label{final-current}
 & & \sigma_{2xy} = \frac{\pi e^3 \tau^2 B_x}{96\hbar^2 (m^*\alpha)}\left\{t-\frac{1}{8}\left[\frac{\lambda (m^*\alpha)^2}{\alpha}\right]^2(1- t)^3\right.\nonumber\\
 &&\left.\times\left\{40 + 192t - 40 t^2\right.\right.\nonumber\\
&& \left.\left. +\left(1-t\right)\left[20 + 35t+20t^2\right]-7\left(1-t\right)^2t\right\}\right\}
 \end{eqnarray}
 Note that this expression is valid for all values of the interpolation parameter $t$, in the one-conduction band regime, as is the case for the surface state of a TI.

In Eq.~(\ref{final-current}) there is a term that vanishes for the warping parameter $\lambda = 0$. However, the nonlinear planar Hall current exists even in the absence of hexagonal warping $\lambda$, as long as the effective mass remains finite ($\frac{1}{m^*\alpha} > 0$). For $\lambda = 0$, we obtain,
\begin{equation}
    \sigma_{2xy} = \frac{\pi e^3 \tau^2 B_x}{96\hbar^2 (m^*\alpha)}t = \frac{\pi e^3 \tau^2 E_x^2 B_x}{96\hbar^2 }\frac{1}{\sqrt{(m^*\alpha)^2 + 2m^*\epsilon_F}}\;.
\end{equation}

The Dirac limit of the topological insulator (TI) is realized when $m^*\alpha^2\gg 2\epsilon_F$, or equivalently $t \simeq 1-\frac{\epsilon_F}{(m^*\alpha^2)}$. Therefore,
\begin{eqnarray}
 && \sigma_{2xy} = \frac{\pi e^3\tau^2  B_x}{96\hbar^2 }\left\{\frac{1}{m^*\alpha} - {24}\frac{\lambda^2}{\alpha^2}\left(\frac{\epsilon_F}{\alpha}\right)^3\right.\nonumber\\
  &&\left.\times\left[1
  +\frac{25}{64}\frac{\epsilon_F}{m^*\alpha}-\frac{7}{192}\left(\frac{\epsilon_F}{m^*\alpha^2}\right)^2\right] \right\}\;. \label{eq:TI-current}
\end{eqnarray}
When $m^*\rightarrow \infty$, the transverse current becomes,
\begin{equation}
\sigma_{2xy} = -\frac{\pi e^3\tau^2 B_x}{4\hbar^2}\frac{\lambda^2}{\alpha^2}\left(\frac{\epsilon_F}{\alpha}\right)^3\;. \label{eq:TI-current*}
\end{equation}
 a result that up to a multiplicative factor is same as that of Ref.~\cite{He_PRL_2019,zhang}.
 Otherwise, for a finite $m^*$, the current is determined by two terms of opposite signs, one proportional to $1/m^*\alpha$ and the other proportional to $\lambda^2$, its overall magnitude and direction being dependent on the exact value of $m^*$, $\epsilon_F$, $\alpha$, and $\lambda$.
 For intermediate values of the interpolation parameter $t$, which depends on the experimental system being investigated, the current can be evaluated from the complete expression for the one-band non-linear Hall current given as a function of $t$ in Eq.~(\ref{eq:TI-current})}.

With $\sigma_1 = e^2\tau\epsilon_F/4\pi\hbar^2$ the linear conductivity of the TI system, one can calculate $\gamma_{yH}$ from Eq.~(\ref{eq:gamma-y}), or alternatively, the second harmonic resistivity
\begin{equation}
\rho_{xy} = A\gamma_{yH} E_xB_x = - \frac{\sigma_{2xy}}{(\sigma_1)^2}E_x = (\chi '-\chi'')E_xB_x\;, \label{eq:rho1}
\end{equation}
where, from Eq.~(\ref{eq:TI-current}),
\begin{eqnarray}
\chi' & = & \frac{4\pi^3\hbar^4}{e}\frac{\lambda^2}{\alpha^5}\epsilon_F\left[1
  +\frac{25}{64}\frac{\epsilon_F}{m^*\alpha}-\frac{7}{192}\left(\frac{\epsilon_F}{m^*\alpha^2}\right)^2\right]\;, \nonumber\\
  \label{chi1-w}
  \end{eqnarray}
  is the contribution of the warping interaction, while
  \begin{equation}
  \chi'' = \frac{\pi^3\hbar^4}{6e}\frac{1}{m^*\alpha\epsilon_F^2}\;,
  \end{equation}
  is the independent contribution of the effective spin-orbit coupling.

  In contrast to the result of Ref.~\cite{He_PRL_2019}, we find that the second-order planar Hall resistivity is determined by the difference $(\chi'-\chi'')$, rather than by their sum.

\subsection{2D surface state Hamiltonian in the 2DEG limit}

For the interpolation parameter $t\rightarrow 0$ ($2m^*\epsilon_F\gg (m^*\alpha)^2$, see Eq.~(\ref{eq:t})) the system resembles a 2DEG with an effective spin-orbit coupling given by the Dirac velocity $\alpha$. In this case, as discussed below Eq.~(\ref{pxi2}), both bands need to be considered in the calculation, along with their corresponding relaxation rates $\tau^2_\xi = \tau^2(1-\xi m^*\alpha/p_0) = \tau^2(1-\xi t)$ from Eq.~(\ref{eq:tau-xi}). In terms of the interpolation parameter $t$ the nonlinear Hall current can be written as,
 \begin{eqnarray}\label{eq:2deg}
 & & \sigma_{2xy} = \frac{\pi e^3 \tau^2 B_x  }{96\hbar^2 (m^*\alpha)}\sum_{\xi = \pm 1}^{}\xi(1-\xi t)\left\{t-\frac{1}{8}\left[\frac{\lambda (m^*\alpha)^2}{\alpha}\right]^2\right.\nonumber\\
 &&\times\left.(1-\xi t)^3\left\{40 + 192\xi t - 40 t^2\right.\right.\nonumber\\
&& \left.\left. +\left(1-\xi t\right)\left[20 + 35\xi t+20t^2\right]-7\xi\left(1-\xi t\right)^2t\right\}\right\}
 \end{eqnarray}

First, we note that in the absence of the warping interaction,
\begin{equation}
\sigma_{2xy} = -\frac{\pi e^3 \tau^2E_x^2 B_x }{48\hbar^2}\frac{m^*\alpha}{(m^*\alpha)^2 + 2m^*\epsilon_F} \;.
\end{equation}
This is the same result one obtains for a 2D Rashba system with a Fermi energy above the band crossing point \cite{Marinescu_2023}.

{In the 2DEG limit, when $t\rightarrow 0$,} the current expression becomes, from Eq.~(\ref{eq:2deg}),
 \begin{eqnarray}
 & & \sigma_{2xy} = -\frac{\pi e^3 \tau^2 E_x^2 B_x }{48\hbar^2m^*\alpha}\left[t^2 + 50 t\left[\frac{\lambda (m^*\alpha)^2}{\alpha}\right]^2\right]\nonumber\\
 && = -\frac{\pi e^3 \tau^2 E_x^2 B_x }{48\hbar^2\sqrt{(m^*\alpha)^2 + 2m^*\epsilon_F}}\left\{\frac{m^*\alpha}{\sqrt{(m^*\alpha)^2 + 2m^*\epsilon_F}}\right.\nonumber\\
 &&\left. + 50 \left[\frac{\lambda (m^*\alpha)^2}{\alpha}\right]^2\right\}\;.
 \end{eqnarray}
Using linear conductivity of the 2DEG system, $\sigma_1 = e^2\tau\epsilon_F/\pi\hbar^2 $, second harmonic resistivity in Eq.~(\ref{eq:rho1}) can be written as,
\begin{equation}
\rho_{xy} = \left(\chi' + \chi''\right)\;,\label{eq:rho2}
\end{equation}
with
\begin{equation}
\chi' = \frac{25\pi^3\hbar^4}{24e\epsilon_F^2}\frac{1}{\sqrt{(m^*\alpha)^2 + 2m^*\epsilon_F}}\left[\frac{\lambda (m^*\alpha)^2}{\alpha}\right]^2\;,
\end{equation}
and
\begin{equation}
\chi'' = \frac{\pi^3\hbar^4}{48e\epsilon_F^2}\frac{m^*\alpha}{(m^*\alpha)^2 + 2m^*\epsilon_F}\;.
\end{equation}

\section{Second order longitudinal magneto-resistance}

To complete this analysis, we calculate the second-order magnetochiral anisotropy-induced magnetoresistance along the $\hat{y}$ direction. In this case, while the direction of the magnetic field remains fixed along $\hat{x}$, the electric field is applied along $\hat{y}$ and the change in resistance is measured parallel to the applied electric field. For this problem, the second order distribution function is given by Eq.~(\ref{eq:ef2}) written for a velocity $v_y$ and an electric field $E_y$. Correspondingly, the longitudinal current is obtained from Eq.~(\ref{current-0}) written for the new distribution function, leading to
\begin{eqnarray}\label{eq:jy*y}
& & j_{yy}^{(2)} = -\frac{e^3 E_y^2}{48\hbar^2}\sum_\xi\tau_\xi^2\int_0^{2\pi}\frac{d}{d\epsilon}\left[v_{y\xi}^3 p_\xi\frac{dp_\xi}{d\epsilon} \right]_{\epsilon_F}d\varphi\;.\nonumber\\
 \end{eqnarray}
In contrast to Eq.~(\ref{eq:K}), the angular kernel of the current integral is now of the form,
\begin{equation}\label{eq:K-y}
\bar{\cal{K}}_\xi = \int_{0}^{2\pi}v_y^3p_\xi\frac{dp_\xi}{d\epsilon}d\varphi\;,
\end{equation}
When linearized in $B_x$, after a long, but otherwise straightforward calculation, we obtain,
\begin{eqnarray}\label{eq:angular-kernel-y}
  \bar{\cal{K}}_\xi &=& \frac{3\pi}{2}B_x\left[-\xi\left(\frac{p_0}{m^*}\right) -\xi\frac{1}{4} p \left(\frac{p_0}{m^*}\right)^2 W_1' +\frac{1 }{2}\left(\frac{p}{m^*}\right) W_0'\right.\nonumber\\
  &+& \left.\frac{1}{2}\left(\alpha - 3\xi\frac{p_0}{m^*}\right)W_1
 +\frac{W_0}{2 p_0} + \xi \frac{W_0}{2m^*\alpha}\right]\;.
\end{eqnarray}
The exact proportionality between $\bar{K}_\xi$ in Eq.~(\ref{eq:angular-kernel}) and $\bar{\cal{K}}_\xi$ given above, the latter being three times as big as the former, is due to the fact that the Taylor expansion in $B_x$ of the velocity driving the current, in this case $v_y$, has a coefficient $3$ in $\bar{\cal{K}}_\xi$ and $1$ in $\bar{K}_\xi$.
This proportionality is carried over in the current calculation, as well as in the expression of the second-harmonic resistivity $\rho_{yy}$. The dependence of the longitudinal current and $\rho_{yy}$ on the interpolation parameters remains unchanged. Thus, $\rho_{yy} = 3\rho_{xy}$ for all values of the interpolation parameter $t$ and, in particular, in both the Dirac fermion and 2DEG with effective spin-orbit coupling regimes of the Hamltonian in Eq.~(\ref{rashba-2D}).

\section{Summary and Conclusions}
 We calculate the nonlinear planar Hall effect in topological insulator surface states with a finite mass $m^*$ described by the Hamiltonian in Eq.~(\ref{rashba-2D}). The first term in $H_{TI}$  proportional to the inverse effective mass breaks particle-hole and sublattice symmetries \cite{Schnyder} and is always present in the surface states of topological insulators \cite{Phillips, Fu_Warping, Liu, McKenzie_2013}.  We define an interpolation parameter $t\in[0,1]$ in Eq.~(\ref{eq:t}), that interpolates the Hamiltonian in Eq.~(\ref{rashba-2D}) between the Dirac fermion limit (large $m^*$, $(m^*\alpha)^2\gg 2m^*\epsilon_F$, $t\rightarrow 1$) and the limit of a 2DEG with an effective spin-orbit coupling given by the Dirac velocity $\alpha$ (small $\alpha$, $(m^*\alpha)^2\ll 2m^*\epsilon_F$, $t\rightarrow 0$). For an intermediate value of $0<t<1$ the Hamiltonian describes a 2D electron gas with an effective spin-orbit coupling $\alpha$. Generically, for this system, the Hamiltonian in Eq.~(\ref{rashba-2D}) allows a pair of Fermi surfaces with opposite chirality index $\xi=\pm 1$. However, the condition on the effective mass $m^*\gtrsim \frac{\Lambda}{2\alpha}$, where $\Lambda$ is a system-dependent wavenumber cutoff above which the Hamiltonian in Eq.~(\ref{rashba-2D}) is no longer valid, effectively eliminates the $\xi = -1$ chiral band from transport considerations. As we have discussed in this paper, this condition can be naturally satisfied for topological insulators with values of the interpolation parameter $t \sim 1$. To preserve this condition of single-band transport for a general value of the interpolation parameter $0< t <1$, one needs to use the constraint that $\epsilon_F\gg (m^*\alpha^2)/2$ for the same values of the effective mass and $\alpha$. For the interpolation parameter $t\sim 0$, the Fermi momenta are close to each other and the system will generically exhibit transport arising from both Fermi surfaces.

 With the above caveats on the Hamlitonian in Eq.~(\ref{rashba-2D}), we show using Boltzmann transport formalism in the nonlinear regime, that in the presence of a parallel configuration of magnetic field $B_x$ and an electric field $E_x$, a current proportional with $E_x^2B_x$ is obtained in the transverse in-plane direction (along the $y$-axis) for all values of $t$ even in the absence of the warping interaction. This is the magnetochiral anisotropy-induced nonlinear planar Hall effect for topological insulator surface states observed in Ref.~[\onlinecite{He_PRL_2019}]. Our principal result is given in Eq.~(\ref{eq:jy-t}), which is a general expression for the second-order planar Hall current in a 2D system with Hamiltonian given in Eq.~(\ref{rashba-2D}) for an arbitrary value of the interpolation parameter $t$. In contrast to the previous literature \cite{He_PRL_2019,zhang} where only the Dirac fermion limit ($t\rightarrow 1$) is discussed this is a significant theoretical improvement whence the nonlinear PHE can be read-off for any value of the interpolation parameter $t$.
 We apply our general result for the nonlinear PHE for arbitrary values of $t$ to the two limiting cases: Dirac fermion limit with linear dispersion relation and single band transport ($t\rightarrow 1$) and the 2DEG limit with an effective SOC $\alpha$ and two-band transport ($t\rightarrow 0$). We find that, for $t\rightarrow 1$, the second-order nonlinear planar Hall resistivity is obtained as the difference of two contributions: one induced by the spin-orbit interaction and the other induced by the warping term, as given in Eq.~(\ref{eq:rho1}). Note that, even though the resistivity $\rho_{xy}$ is linear in $E_x$ and $B_x$, the nonlinear planar Hall current is proportional to $E_x^2$ and $B_x$ (see Eq.~(\ref{eq:jy-t}). Depending on the values of the microscopic parameters of the sample, one of the two currents will dominate the nonlinear planar Hall effect. In the 2DEG limit ($t\rightarrow 0$), both chiral bands need to be included in the calculation of the nonlinear planar Hall effect. This regime of the surface state Hamiltonian has not been discussed before, except in the context of simple Rashba systems \cite{He_PRL_2019} where the non-linear planar Hall effect was found to vanish due to the cancellation between the two chiral bands. In this paper, we confirm that the nonlinear PHE is nonzero even in this limit ($t \rightarrow 0$) and the corresponding expression for the resistivity is given in Eq.~(\ref{eq:rho2}) where the two currents add. In addition, we also calculate the second-order magnetochiral anisotropy-induced magnetoresistance in a geometry in which the electric and the in-plane magnetic fields are perpendicular to each other and the change in resistance is measured parallel to the electric field. In this geometry, we show that a nonlinear magnetoresistance is possible in TI surface states in which the additional longitudinal current is proportional to the magnetic field and two powers of the electric field and is three times in magnitude of the nonlinear planar Hall current discussed in this paper.

For the sample characteristics given in Ref.~\cite{He_PRL_2019}, carrier
concentration $n \sim 7 \times 10^{13}$ cm$^{-2}$, Fermi energy $\epsilon_F\sim
400$ meV,  $\alpha = 5\times 10^5$m/s, and $m^* = 0.07$, we note that $m^*\alpha^2/2 \sim 50$meV, leading to a parameter $t = 0.33$. For this value, the single-band Dirac fermion limit is not valid.
Based on our theory, the current that was measured in the experiment discussed in Ref.~\cite{He_PRL_2019} is associated with the two-band transport regime, in which the spin-orbit and warping contributions add as in Eq.~(\ref{eq:rho2}). Our results given in Eq.~(\ref{eq:jy-t}) are more generally valid, however, and experiments of nonlinear PHE for arbitrary values of $t$ can be compared with Eqs.~(\ref{eq:jy-t}, \ref{eq:rho1}, \ref{eq:rho2}) depending on the values of the sample-specific interpolation parameter given in Eq.~(\ref{eq:t}).

\section{Acknowledgement}
ST acknowledges
support from the Army Research Office through Grant No: W911NF2210247 and the Office of Naval Research through Grant No: N00014-23-1-2061.

\end{document}